\documentclass[sigconf, anonymous=false]{acmart}

\usepackage{xspace}
\usepackage{xfrac}
\usepackage{subcaption}
\usepackage{multirow}
\usepackage{enumitem}
\usepackage{listings}
\usepackage{xcolor}
\usepackage{float}

\definecolor{codebg}{RGB}{245,247,250}   
\definecolor{codeframe}{RGB}{220,220,220}

\lstdefinestyle{SASS}{
  language=[LaTeX]Tex, 
  basicstyle=\ttfamily\footnotesize, 
  keywordstyle=\color{blue}\bfseries, 
  commentstyle=\color{gray}\itshape,
  stringstyle=\color{red},
  breaklines=true,
  numbers=left,
  numberstyle=\tiny\color{gray},
  stepnumber=1,
  numbersep=5pt,
  showstringspaces=false,
  backgroundcolor=\color{codebg},
  frame=single,
  rulecolor=\color{codeframe},  
  captionpos=b
}

 \AtBeginDocument{%
  }

\setcopyright{acmlicensed}
\copyrightyear{2026}
\acmYear{2026}
\setcopyright{cc}
\setcctype{by}
\acmConference[ICS '26]{2026 International Conference on Supercomputing}{July 06--09, 2026}{Belfast, United Kingdom}
\acmBooktitle{2026 International Conference on Supercomputing (ICS '26), July 06--09, 2026, Belfast, United Kingdom}
\acmDOI{10.1145/3797905.3807868}
\acmISBN{979-8-4007-2522-7/2026/07}

\begin{document}

\title{\textsc{Ocean}: Fast Estimation-Based Sparse General Matrix-Matrix Multiplication on GPU} 

\author{Yifan Li}
\affiliation{%
  \institution{Cornell University}
  \city{Ithaca, NY}
  \country{USA}}
\email{yl3722@cornell.edu}

\author{Giulia Guidi}
\affiliation{%
  \institution{Cornell University}
  \city{Ithaca, NY}
  \country{USA}
}
\email{gg434@cornell.edu}

\renewcommand{\shortauthors}{Li and Guidi}

\begin{abstract}
In computational science and data analytics, many workloads involve irregular and sparse computations that are inherently difficult to optimize for modern hardware. A key kernel is Sparse General Matrix--Matrix Multiplication (SpGEMM), which underpins simulations, graph analytics, and machine learning applications. SpGEMM exhibits irregular memory access patterns and workload imbalance, making it challenging to achieve high performance on GPUs.
Current GPU SpGEMM solutions typically rely on a two-pass workflow to address load imbalance and reduce memory access. The symbolic pass, which determines the number of output elements per row, accounts for roughly 28\% of the total runtime on average. 
In this work, we question the necessity of exact symbolic computation and introduce an estimation-based SpGEMM workflow.

Our approach replaces the costly symbolic step with lightweight HyperLogLog estimators, combined with an analysis strategy that dynamically selects the optimal workflow and guides accumulator configuration. 
In addition, we introduce a hybrid accumulator design, including a novel hash-based accumulator that leverages both shared and global memory.

Our approach consistently outperforms leading GPU SpGEMM implementations across a wide range of both square and rectangular matrices, achieving speedups of $1.4\times$–$2.8\times$ on NVIDIA A100 and H100 architectures.

\end{abstract}

\ccsdesc[500]{Theory of computation~Massively parallel algorithms}
\ccsdesc[500]{Computing methodologies~Linear algebra algorithms}



\newcommand{\ours}{\textsc{Ocean}\xspace}
\newcommand{\GG}[1] {{{\color{red}GG: #1}}}
\newcommand{\YL}[1] {{{\color{blue}YL: #1}}}

\newcommand{\B}{\mathbf{B}\xspace}
\newcommand{\A}{\mathbf{A}\xspace}
\newcommand{\C}{\mathbf{C}\xspace}

\maketitle

\section{Introduction}

Data-driven, irregular, and sparse computations are central to modern data analytics and scientific computing, but they are notoriously hard to optimize on current hardware. 
A key kernel is Sparse General Matrix--Matrix Multiplication (SpGEMM), which underpins simulations, graph analytics, and machine learning applications. 
SpGEMM multiplies two sparse matrices, $\A$ and $\B$, to produce an output matrix $\C$. 
This operation is fundamental to many applications, including algebraic multigrid, Markov clustering, and computational biology~\cite{guidi2021bella, guidi2021parallel, selvitopi2020distributed, azad2018hipmcl, buluç2025ubiquitoussparsematrixmatrixproducts}. 
It is also used in machine learning, such as in graph neural networks~\cite{wang2019dgl, tripathy2020reducing}, and has the potential to be adopted in sparse attention mechanisms. 
GPUs have recently become the focus of SpGEMM research, as their high computational throughput and memory bandwidth make them more powerful than CPUs, especially for large matrices.

The underlying mathematical operation is the same as general matrix multiplication, but the high sparsity of the input matrices makes conventional dense matrix multiplication inefficient. 
SpGEMM implementations~\cite{parger_speck_2020,du_opsparse_2022,wu_hsmu-spgemm_2025,wang_optimizing_2025,islam_improving_2025,winter_adaptive_2019} operate only on the nonzero elements of the input matrices to avoid unnecessary computation on zero entries.
This computation paradigm, however, poses challenges for parallelization, especially on massively parallel architectures such as GPUs.
The often irregular nonzero patterns of input matrices result in unpredictable computation and memory access behavior. 
Moreover, the nonzero pattern of the output matrix and the number of intermediate products are unknown before execution. 
Therefore, efficient approaches depend on dynamic load balancing and fine-grained allocation of on-chip resources.

Currently, most GPU-based SpGEMM solutions use variants of Gustavson's algorithm to address these issues. 
Gustavson's algorithm computes each row of the output matrix $\C$ independently, providing natural parallelization opportunities on GPUs, where a thread block is typically assigned to each output row. 
This reduces the scratchpad memory required per block and allows load balancing across rows to be partially managed by the hardware scheduler.
In addition, state-of-the-art solutions typically adopt a two-pass approach to fully utilize hardware resources during numeric computation~\cite{parger_speck_2020,du_opsparse_2022,wu_hsmu-spgemm_2025}.
The symbolic pass is executed first to calculate the number of nonzero elements in each output row. 
The numeric pass then performs the numerical computations and writes the results. 
The output of the symbolic pass is critical for guiding the allocation of scratchpad memory during the numeric pass to ensure efficient resource utilization.

Both the symbolic pass and numeric pass use accumulators to store temporary results. Different types of accumulators, such as hash-based and dense accumulators, have been studied extensively \cite{parger_speck_2020, du_opsparse_2022, wu_hsmu-spgemm_2025, liu2014efficient, winter_adaptive_2019, gilbert1992sparse, dalton2015optimizing}. 
Each accumulator offers advantages for specific matrix types. The accumulator is stored in scratchpad memory, with global memory used as a fallback when the required capacity exceeds the scratchpad's maximum limit~\cite{parger_speck_2020, du_opsparse_2022, cusparse}.

Current strategies have addressed several major challenges in GPU-based SpGEMM, including dynamic accumulator selection~\cite{parger_speck_2020}, global and local load balancing~\cite{winter_adaptive_2019,parger_speck_2020}, and architecture-specific optimizations~\cite{du_opsparse_2022,wang_optimizing_2025}. 
However, computational efficiency remains well below hardware limits. 
For example, \textsc{spECK} and \textsc{HSMU-SpGEMM} achieve only about $3-4\%$ of peak FP64 performance on an RTX 3090 Ti on average~\cite{wu_hsmu-spgemm_2025}, largely due to memory access bottleneck and complex control flow.
The two computation passes, often exceeding $85\%$ of total runtime, constitute the main bottleneck.

\vspace{-.5em}
\paragraph{\textbf{Challenges}}
Two notable limitations in the widely adopted computational paradigm for GPU-based SpGEMM can restrict its efficiency.
\textcircled{\small 1} The cost of size prediction can outweigh the benefits of load balancing and precise resource allocation. 
Our experimental results show that, for the state-of-the-art solution \textsc{spECK}~\cite{parger_speck_2020}, the symbolic step accounts for $\approx28\%$ of the total execution time, even though it produces only a single value per output row.
A lightweight solution for size prediction has the potential to significantly reduce this overhead.
Current output size estimation approaches often lack accuracy and granularity, and can be more costly than symbolic computation for some matrices.
\textcircled{\small 2} Current accumulator designs may not perform efficiently for both short and long output rows. 
Short rows, defined as having fewer nonzeros than a warp (typically 32), underutilize hardware due to their limited size.
In contrast, long rows, with more nonzeros than the capacity of a hash accumulator on a single streaming multiprocessor, revert to non-adaptive global-memory kernels and incur substantial performance penalties on large matrices.

\vspace{-.5em}
\paragraph{\textbf{Contributions}}
In this paper, we address these challenges and propose an estimation-based SpGEMM solution. 
In particular, we replace the symbolic pass with fast HyperLogLog-based estimation. 
HyperLogLog~\cite{flajolet_hyperloglog_2007} provides efficiency and sufficient accuracy for per-row output size prediction. 
Because the estimation-based workflow may incur higher computation and memory access costs, we introduce an analysis step that uses input statistics and sampling to dynamically select between estimation-based and symbolic-based workflows. 
The information gathered during this step is also used to further optimize the accumulation kernels.
In addition, we optimize the accumulation of short and long output rows, often overlooked by existing solutions, through a hybrid accumulator design and cooperation between shared and global memory.

Our solution, \ours, is implemented in CUDA C++\footnote{Code available at \url{https://github.com/CornellHPC/Ocean-SpGEMM}}.
Our experiments, conducted on an NVIDIA A100 platform using 337 square matrices, show that \ours achieves speedups of $1.4\times$, $2.6\times$, $3.5\times$, and $2.0\times$ over \textsc{spECK}, \textsc{opSparse}, \textsc{TileSpGEMM}, and \textsc{HSMU-SpGEMM}, respectively. 
Consistent speedup is also observed on the NVIDIA H100 platform and on a dataset of 64 rectangular matrices.
In summary, our main contributions include:
\begin{enumerate}[leftmargin=*,noitemsep]

\item A new paradigm for GPU SpGEMM that uses HyperLogLog estimation instead of exact symbolic computation. 
To the best of our knowledge, this is the first work to use HyperLogLog to accelerate sparse linear algebra, highlighting the potential of estimators for sparse primitives;

\item An analysis step that accurately predicts the cost of estimation and selects the best workflow with minimal overhead;

\item A hybrid accumulator design that leverages three types of accumulators and a specialized configuration that exploits cooperation between shared and global memory;

\item \ours, an open-source GPU implementation, that achieved speedups of $1.4\times$ to $2.8\times$ over the state-of-the-art.

\end{enumerate}

\section{Background}

In this section, we provide background information on the basics of SpGEMM, state-of-the-art approaches for SpGEMM on GPUs, and HyperLogLog estimator.

\subsection{CSR format and Gustavson’s Algorithm}

Given sparse matrices $\A \in \mathbb{R}^{m \times k}$ and $\B \in \mathbb{R}^{k \times n}$, sparse matrix– matrix multiplication computes $\C = \A\B$, where $\C \in \mathbb{R}^{m \times n}$.
For SpGEMM computation, the input matrices are typically stored in Compressed Sparse Row (CSR) format, where two arrays store the column indices and values of the nonzero entries. 
The nonzeros in each row are stored contiguously, and a sorted array of row offsets indicates the starting position of each row. 
This storage format reduces space usage by eliminating redundant row index storage.

The CSR format is particularly well suited for SpGEMM computation using Gustavson's algorithm~\cite{gustavson1978two}. 
Gustavson's algorithm computes each row of the output matrix $\C$ independently, with each row depending only on a single row of the input matrix $\A$.
Each row is computed using two nested loops.
The outer loop traverses the nonzeros in the corresponding row of $\A$, and the inner loop traverses the row of $\B$ associated with each of these nonzeros.
During this accumulation process, intermediate results are stored in memory using accumulators. 
Because Gustavson's algorithm accesses nonzeros within the same row of both input matrices contiguously, the CSR representation enables efficient traversal. 
The combination of the CSR format and Gustavson's algorithm is used in most SpGEMM approaches~\cite{liu2014efficient, nagasaka_high-performance_2017, parger_speck_2020, du_opsparse_2022, wu_hsmu-spgemm_2025, cusparse}.

\begin{figure}[t]
  \centering
  \includegraphics[width=1.0\columnwidth]{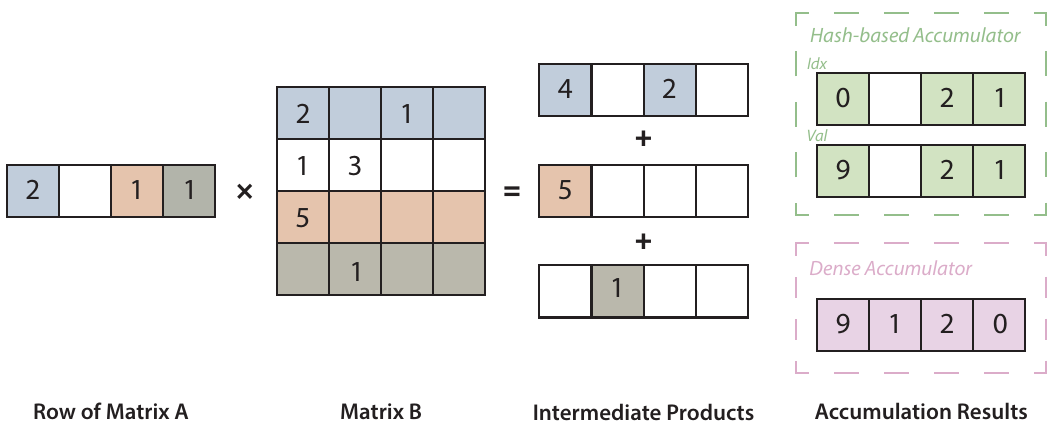}
  \caption{
  Computation of a single output row using Gustavson’s algorithm with a hash-based and dense accumulator.
    Nonzero elements of the input row are multiplied with their corresponding rows in matrix $\B$ to generate intermediate products. \
    These intermediate products are stored in accumulators until all computations are completed. \
  }
  \label{fig:gustavson}
\end{figure}

\subsection{Accumulators}

Gustavson’s algorithm uses accumulators to store intermediate results until an entire output row is computed. 
Figure~\ref{fig:gustavson} illustrates a simple example of computing a single output row using Gustavson’s algorithm with two different accumulator designs. 
GPUs consist of multiple independent compute units called \textit{streaming multiprocessors} (SMs). 
Each SM has a fixed amount of hardware resources, including compute cores, registers, and fast on-chip scratchpad memory. 
Scratchpad memory not explicitly allocated to shared memory is automatically configured as L1 cache, while SMs share a unified L2 cache. 
Global memory is shared by SMs and is much larger but significantly slower. 
Because accumulators are accessed frequently, they are typically placed in scratchpad memory. 

The simplest design is the \textit{dense accumulator}, in which the entire output row is represented as a dense array~\cite{gustavson1978two}. 
This approach provides fast access but is impractical for many matrices because the output row may exceed the limited capacity of scratchpad memory.

\textit{Hash-based accumulators} are more space-efficient and are commonly used in GPU SpGEMM implementations~\cite{gao_systematic_2023,parger_speck_2020,du_opsparse_2022,cusparse}. 
They use a hash table to store the column indices and values of nonzero output elements. 
During numeric computation, the hash table size is typically set to $\sim1.5\times$ the number of nonzero elements in the output row to reduce hash collisions~\cite{parger_speck_2020, dalton2015optimizing, du_opsparse_2022}.
Compared to dense accumulators, hash-based designs are generally more memory-efficient because of the sparsity of output rows, even though they require storing element indices.
Their access latency is higher, as each update involves at least one additional lookup for the index and sometimes requires atomic exchanges on the hash table.

\textit{ESC accumulators} (\textit{Expand–Sort–Compact}) use a different strategy~\cite{bell12anh}.
Rather than merging intermediate products with identical column indices during row traversal, they preserve the intermediate products. Once all products are generated, a sorting step groups products with the same column indices. 
Finally, a compaction step aggregates these groups to produce the output row. 
ESC accumulators require more temporary memory and are generally faster for matrices with small numbers of intermediate products.

\subsection{GPU SpGEMM Workflow}

\begin{figure}[t]
  \centering
  \includegraphics[width=0.5\columnwidth]{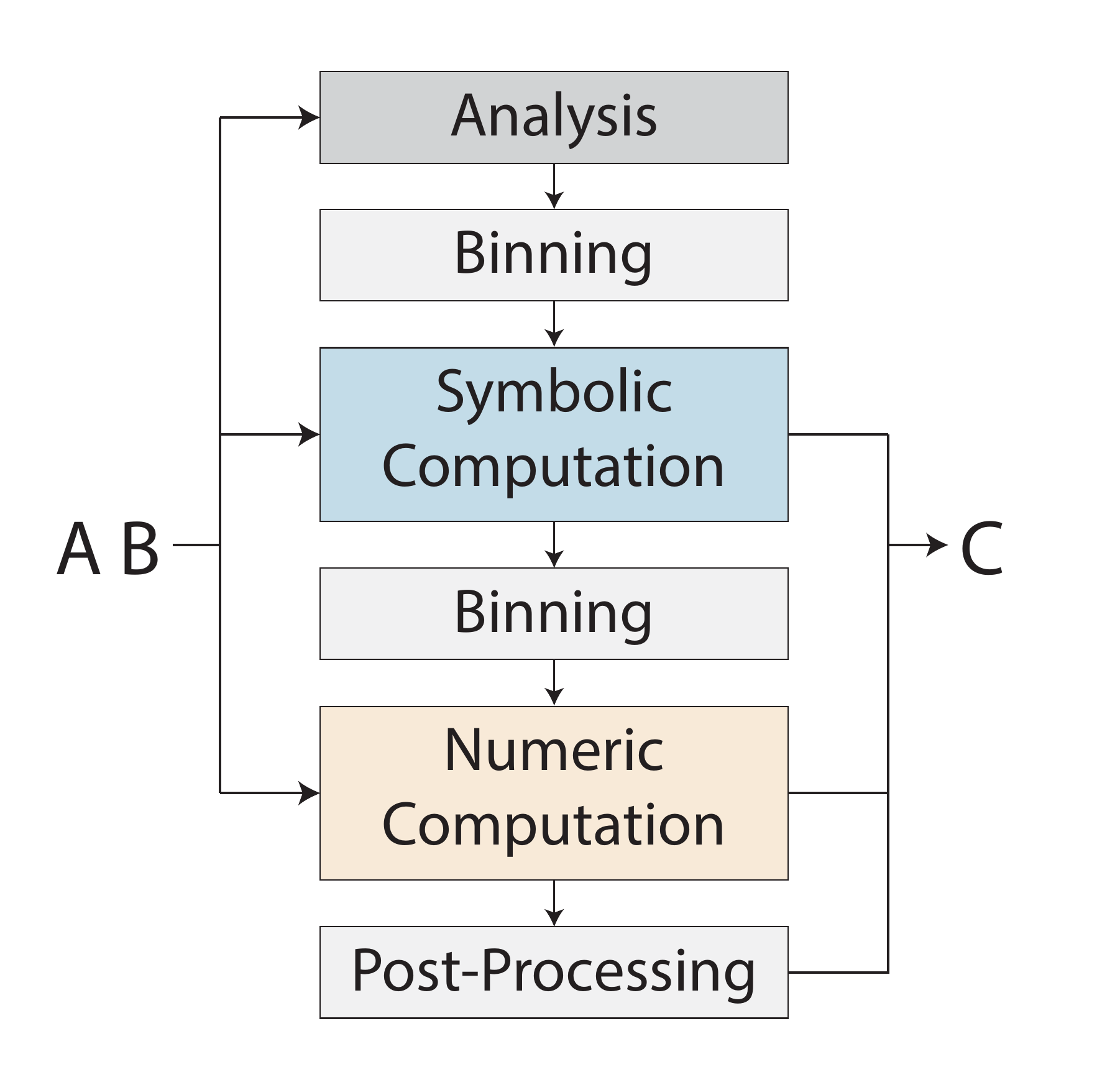}
  \caption{Common GPU SpGEMM Workflow.}
  \label{fig:common_workflow}
  \vspace{-.5em}
\end{figure}

Because they efficiently use scratchpad memory and offer great flexibility, hash-based accumulators are the most commonly used accumulators for GPU SpGEMM~\cite{gao_systematic_2023, parger_speck_2020, du_opsparse_2022, anh2016balanced, nagasaka_high-performance_2017}.
To achieve optimal efficiency, the size of the hash table must be determined for each row. 
However, on GPUs, shared memory is allocated to a thread block before execution. 
As a result, achieving high utilization requires knowing the number of output elements before numeric accumulation begins.
This size estimation phase is commonly called the \textit{symbolic pass}, while the subsequent computation is known as the \textit{numeric pass}. 
Most recent GPU SpGEMM approaches use a symbolic pass for accurate size prediction.

The prediction is also motivated by the need to determine the output matrix structure and minimize memory movement. 
Due to the constraints of the CSR format, output indices and values must be stored in contiguous memory. 
If the exact per-row output size is not known in advance, the computed indices and values cannot be written directly to their final positions in the output CSR matrix. 
As a result, an additional memory copy step is required to reorganize the results into CSR format.

Gustavson's algorithm is also commonly used for symbolic computation. 
During symbolic accumulation, numerical values are discarded, and only the column indices of the output elements are retained. 
This reduces the scratchpad memory requirement by $67\%$ for hash-based accumulators and by $98\%$ for dense accumulators, assuming column indices are stored in \texttt{uint32} and values in \texttt{fp64}. 
Thus, symbolic computation is generally more efficient than numeric computation.

Fine-grained accumulator selection is another key factor in achieving good performance~\cite{parger_speck_2020, du_opsparse_2022, wu_hsmu-spgemm_2025}. 
Because allocating the exact scratchpad memory needed for each row is impractical in GPU programming, a common strategy is to predefine multiple kernel configurations with fixed scratchpad sizes and thread counts.
Rows are assigned to different kernels based on their output size before computation kernels are launched. This process is typically referred to as \textit{binning}.
Certain implementations further extend this approach by supporting multiple types of accumulators~\cite{parger_speck_2020,liu2014efficient}; for example, \textsc{spECK} uses both hash-based and dense accumulators. 
Hash-based accumulators are generally more space-efficient for sparse rows, while dense accumulators provide lower latency and better space efficiency for rows whose nonzero entries are narrowly distributed---that is, the distance between the first and last nonzero is small.
To enable this selection, the binning or computation steps must choose between accumulator types, which requires additional information, such as the nonzero span of each output row. 
For this purpose, some solutions introduce an \textit{analysis step} as the first stage of the workflow to collect relevant data~\cite{nagasaka_high-performance_2017, parger_speck_2020, du_opsparse_2022, wu_hsmu-spgemm_2025}.

Figure~\ref{fig:common_workflow} illustrates the common workflow.
The symbolic step computes the number of output elements for each row, while the numeric step performs the actual computation. 
Binning is applied before both the symbolic and numeric steps to maximize performance.
The numeric binning is guided by the number of output nonzeros per row, and the symbolic binning is typically guided by the number of intermediate products, serving as a safe upper bound. 
A post-processing step finalizes the output to comply with the CSR format (e.g., if the numeric computation produces unsorted rows, the post-processing step sorts the entries in each row).

\subsection{HyperLogLog}

HyperLogLog (HLL) is a probabilistic cardinality estimation algorithm that provides near-optimal space efficiency for counting distinct elements in data streams~\cite{flajolet_hyperloglog_2007}. 
It is based on the observation that the distribution of leading zeros in uniformly hashed values provides information about the number of distinct elements.

Each HLL sketch maintains a compact array of $m$ registers.
Each element is hashed and assigned to a register using a fixed number of lower hash bits. 
Each register records the maximum number of leading zeros observed among the hash values assigned to it. 
The collection of registers captures the distribution of these maxima, reflecting the probability distribution of the dataset's cardinality.
By computing the harmonic mean across the $m$ registers and applying a statistical correction, HLL provides an unbiased estimate with a relative error of $1.04/\sqrt{m}$. 
As the number of registers in the sketch increases, the precision of the HLL estimator improves.

Due to its small memory footprint and high efficiency, HLL has been widely adopted in large-scale data processing for query optimization, network monitoring, and analytics over massive datasets \cite{heule_hyperloglog_2013}.
To the best of our knowledge, it has not been applied to SpGEMM or any sparse linear algebra primitives.
\section{Methods}

In this section, we describe our SpGEMM solution, \ours, which stands for \underline{O}ptimizing SpGEMM with \underline{C}ardinality \underline{E}stimation \underline{an}d Hybrid Accumulator.
\ours optimizes the symbolic and numeric steps, which together take over 85\% of runtime.
\ours uses Gustavson’s method for numeric computation but differs from the traditional two-pass approach by using HyperLogLog to predict per-row output sizes and guide kernel selection.
Our redesigned workflow introduces a lightweight analysis step for cost prediction and workflow selection, along with an overflow handling mechanism, enabling robust performance across diverse matrices. 
Finally, \ours employs specialized numeric accumulators for short and long rows through a hybrid design.

\subsection{Prediction with HyperLogLog}

\begin{figure}[t]
  \centering
  \includegraphics[width=1.0\columnwidth]{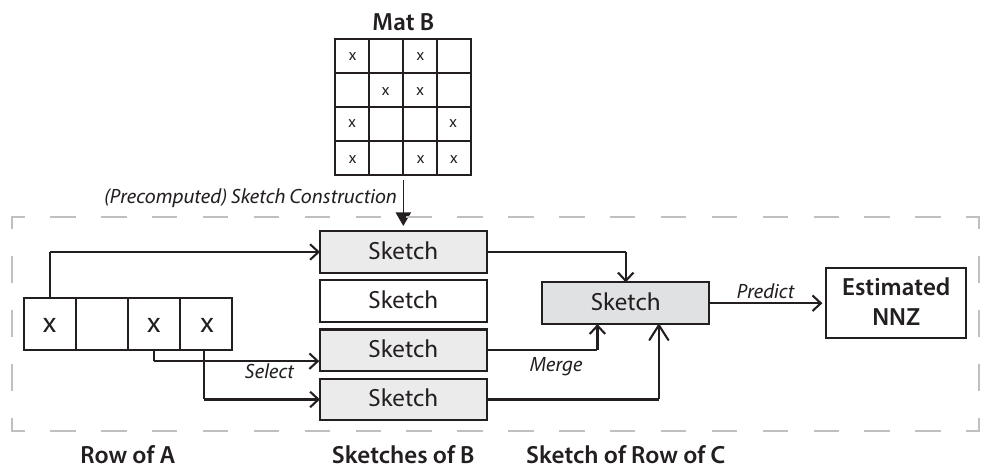}
  \caption{ \
  The estimation of per-row nonzeros using HyperLogLog with construct-and-merge. 
  The sketches are built for each row of matrix $\B$, merged according to each row of matrix $\A$, and the number of output nonzeros are estimated from the merged sketch.
  }
  \label{fig:hll_est}
\end{figure}

HyperLogLog (HLL) is adopted for per-row output size prediction in SpGEMM and is well suited to this task for three key reasons:
\begin{enumerate}[leftmargin=*,noitemsep]
    \item \textbf{Controlled error:} The estimation error remains within an acceptable range with a reasonable number of registers per sketch (see Section~\ref{subsec:eval_prec}).
\item \textbf{Constant memory footprint:} HLL uses a constant amount of memory regardless of the number of nonzeros, making it effective for long rows where traditional accumulators require large kernels or global buffers.
\item \textbf{Parallel updates:} The sketch updates avoid compare-and-swap operations and can be implemented using \texttt{atomicMax}.
\end{enumerate}

These properties allow HLL to serve as a high-performance substitute for the accumulators used in the symbolic step.
A key advantage of HyperLogLog is its ease of merging, which further increases the efficiency of size prediction. 
Combining multiple HLL sketches only requires taking the element-wise maximum across corresponding registers, provided the same hash function is used to construct them.
This property is especially useful for per-row size prediction.
Because the rows of matrix $\B$ are typically accessed multiple times, a sketch can be constructed for each row of $\B$ as a preprocessing step. 
Consequently, for each row of $\A$, the corresponding sketches can be merged to produce the sketch of the corresponding row in $\C$.
The nonzeros in each row of $\C$ can then be estimated from the merged sketch, as shown in Figure~\ref{fig:hll_est}.

This construct-and-merge approach further reduces computation and memory access as long as the number of registers is smaller than the average number of elements per row of $\B$. 
Moreover, the irregular symbolic accumulation is transformed into two regular, cache-friendly stages: sketch construction and sketch merging. 
Overall, HLL-based estimation avoids the high cost of symbolic computation while providing sufficiently accurate per-row size prediction.

\subsection{Estimation-Based Workflow}
\label{subsec:est_wkfl}

\begin{figure}[t]
  \centering
  \includegraphics[width=1.0\columnwidth]{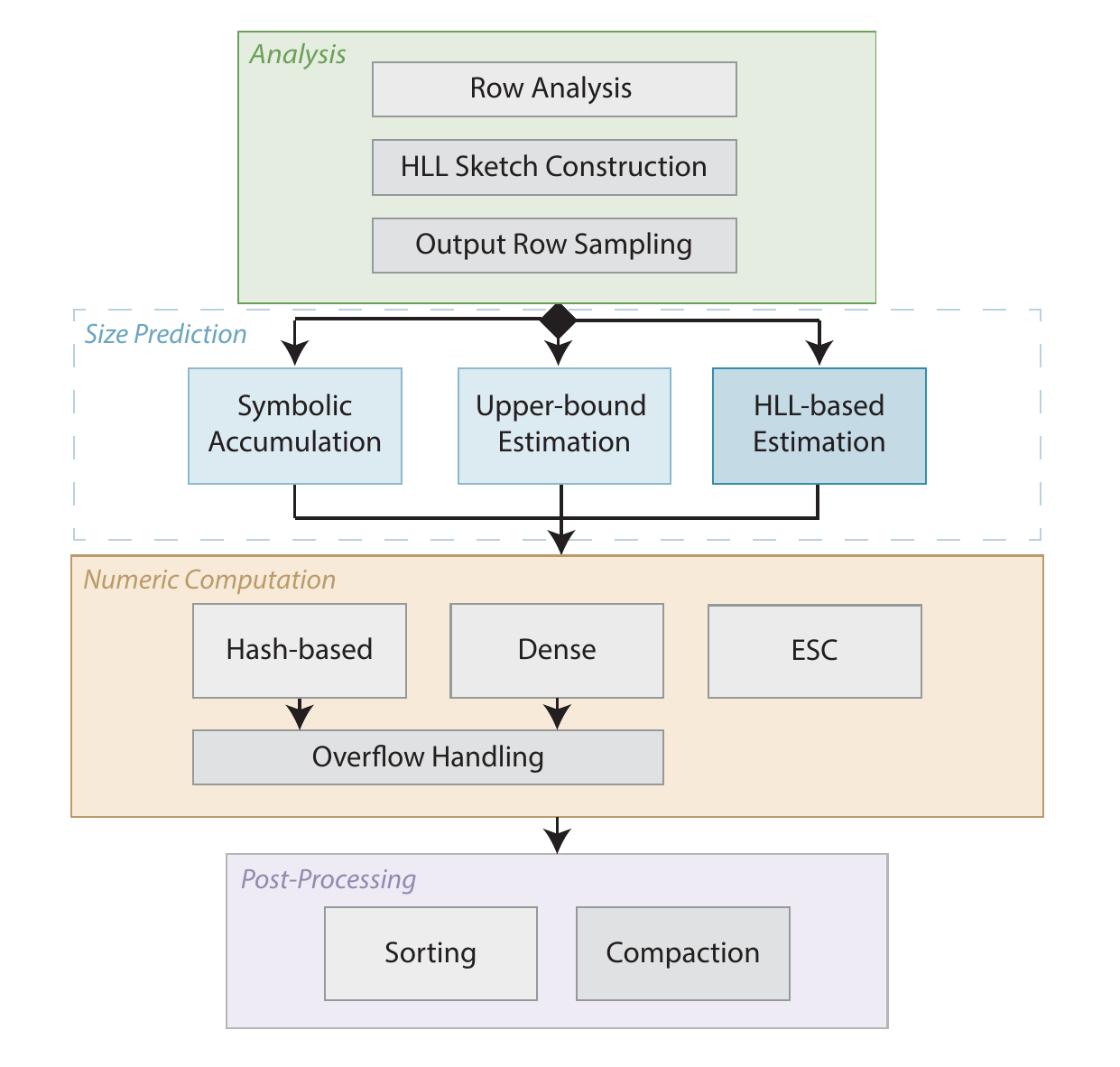}
  \caption{Overview of the \ours SpGEMM workflow, with estimation-based components highlighted in darker backgrounds. The analysis step gathers metrics for workflow selection and load balancing. Then, size prediction uses either symbolic accumulation or estimation to predict per-row output sizes. Finally, numeric computation performs the multiplication using the collected statistics, followed by post-processing to produce the CSR output.
  }
  \label{fig:ocean_workflow}
\end{figure}

HLL-based estimation cannot be integrated into existing SpGEMM solutions by simply replacing the symbolic component. Because the HLL estimator does not guarantee a strict error bound, a fallback mechanism must be provided to handle overflow.
In addition, when matrix $\B$ is highly sparse, the cost of estimation may exceed that of the traditional symbolic step.
In such cases, it is preferable to use the standard symbolic approach. 
In this work, we redesign the SpGEMM workflow and its key components to support estimation-driven execution. Figure~\ref{fig:ocean_workflow} illustrates the resulting \ours pipeline.
The discussion below focuses on the two central challenges of this approach: (i) overflow handling and (ii) cost prediction.

\paragraph{\textbf{Overflow Handling}}
Overflow during numeric accumulation occurs when HLL underestimates the number of output elements in a row, which can severely degrade performance. 
Fortunately, common accumulation kernel designs can tolerate estimation errors within a certain range.
The hash table expansion factor, typically set to $1.5\times$, tolerates some underestimation without noticeable performance impact.
Moreover, the binning process rounds up the expanded hash table size to a predefined bin size, further absorbing estimation errors. 
These two mechanisms greatly reduce the risk of overflow.
However, numeric accumulation may still overflow in rare cases. 
Both hash-based and dense accumulators are susceptible to overflow. 
For hash-based accumulators, accumulation cannot proceed when the allocated shared memory is insufficient. 
For dense accumulators, accumulation proceeds safely because shared memory is allocated according to the output index range. 
Under the current design with pre-allocated global output memory, however, the final results may exceed the available space.

\textsc{Ocean} handles overflow with a single fallback kernel, launched after all normal accumulation kernels complete. 
This kernel uses the largest dense configuration, iterates over the entire row, and can handle arbitrarily long rows. 
Output memory is allocated based on the number of intermediate products, which serves as an upper bound. 
This approach uses conservative resource allocation and therefore reduces efficiency; however, the overall impact is limited because overflow occurs in only a small fraction of rows.

\begin{figure}[t]
  \centering
  \includegraphics[width=1.0\columnwidth]{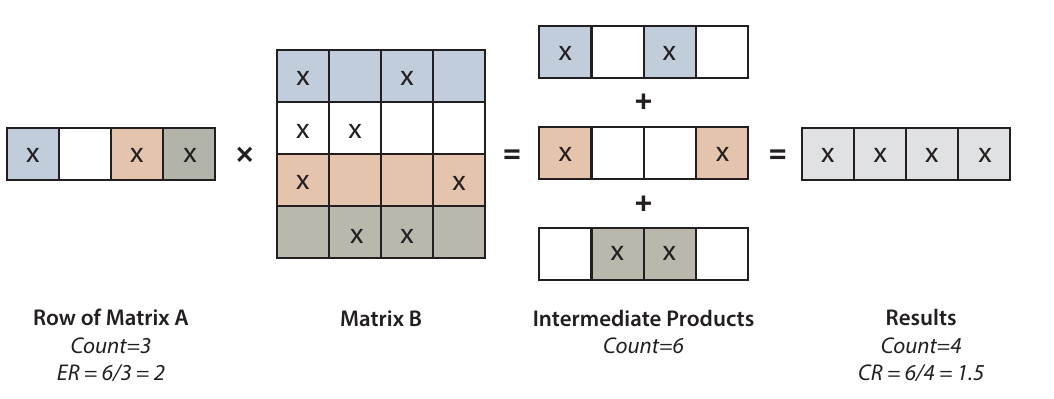}
  \caption{Computation of Input Expansion Ratio ($ER$) and Output Compression Ratio ($CR$). \
$ER$ denotes the ratio between the number of intermediate products and the number of nonzeros in input matrix $\A$.\
$CR$ denotes the ratio between the number of intermediate products and the number of nonzeros in output matrix $\C$.\
  }
  \vspace{-.5em}
  \label{fig:er_cr}
\end{figure}

\paragraph{\textbf{Cost Prediction and Workflow Choice}}
HLL-based estimation is generally efficient, but, in some cases, it can be more expensive than symbolic accumulation. 
An effective yet lightweight analysis step is required to predict the relative cost of estimation and revert to the symbolic approach when necessary. 
Here, we propose the use of the \textit{Input Expansion Ratio (ER)} and \textit{Output Compression Ratio (CR)} for this purpose.

The input expansion ratio is used to estimate the relative cost of the symbolic step compared to the estimation step.
HLL estimators require a fixed number of registers to guarantee a target precision, regardless of the number of processed elements.
For some input matrices, merging HLL sketches requires more computation and memory traffic than a full accumulation pass, making the construct-and-merge approach inefficient. 
Here, we define the Expansion Ratio ($ER$) as the ratio of intermediate products to the number of nonzeros in matrix $\A$.
$ER$ estimates the relative memory traffic of symbolic accumulation compared to HLL-based estimation. 
Because size prediction is typically memory-bound, if this ratio falls below a predefined threshold (e.g., one quarter of the number of registers), the matrix is considered very sparse and the symbolic-based workflow is enabled.
These statistics are collected through an $\mathcal{O}(nnz_\A)$ analysis of matrix $\A$, similar to the analysis step in \textsc{spECK}, where $nnz_\A$ denotes the number of nonzeros in $\A$.

The output Compression Ratio $CR$ has been introduced in prior work~\cite{nagasaka2019performance, parger_speck_2020, du_opsparse_2022} to characterize different matrices.
It is defined as the ratio of the number of intermediate products to the number of nonzeros in the output matrix $\C$. 
In this context, we use $CR$ to estimate the relative cost of the additional post-processing required by the estimation-based workflow. 
Recall that the CSR format requires indices and values to be stored in contiguous memory. 
Unlike symbolic-based approaches that determine the output structure in advance, estimation-based approaches obtain this information only after numeric computation is complete.
A compaction step at the end is required to relocate the output nonzeros and produce a valid CSR representation, which introduces additional memory movement overhead.
$CR$ measures the relative cost between symbolic accumulation and compaction overhead in terms of memory traffic; lower values indicate higher compaction memory traffic and favor a non-estimation-based workflow.

A precise or approximate value of $CR$ serves as a useful indicator, but it cannot be determined until a complete symbolic or estimation step has been performed.
Relying on the estimated $CR$ computed from a full estimation pass would introduce unnecessary overhead for matrices with high compaction costs.
To address this limitation, we introduce the \textit{Sampled Compression Ratio}.
During the analysis step, an HLL sketch is constructed for each row of matrix $\B$.
A subset of rows from matrix $\A$ is randomly selected as samples, and the corresponding HLL sketches are merged and evaluated.
The resulting sampled $CR$ serves as an indicator for selecting either the estimation-based workflow or the precise symbolic step. The cost of computing the sampled $CR$ is kept low to ensure the analysis step remains lightweight.
HLL sketch construction has a complexity of $\mathcal{O}(nnz_\B)$.
Only a small fraction of rows in matrix $\A$ ($3\%$) are sampled and evaluated.
The sketch construction and merging process accounts for only $3\%$ of the total runtime on average.

To further optimize performance for a small number of extremely sparse matrices, both symbolic accumulation and HLL-based estimation are disabled when the average number of intermediate products per row falls below a lower threshold. 
In this case, the estimation-based workflow is selected, using the number of intermediate products as an upper-bound estimate.

Overall, our analysis uses multiple metrics to choose between estimation-based workflows (including upper-bound estimation) and the symbolic-based workflow. From matrices $\A$ and $\B$, we obtain $ER$ and the number of intermediate products per row, while row sampling estimates $CR$. Figure~\ref{fig:er_cr} illustrates these statistics.
Together, these statistics characterize the relative cost of estimation-based workflows and can be computed with minimal overhead. 
The workflow selection criteria based on these statistics are presented in Section~\ref{subsec:param_conf} and Table~\ref{tab:criteria}.

\subsection{Hybrid Accumulator}

Then, we redesign the hybrid accumulator scheme to better handle very short and very long rows.
Prior hybrid design has combined hash-based and dense accumulators with multiple kernel configurations~\cite{parger_speck_2020}.
Building on this approach, we introduce:
\begin{enumerate}[leftmargin=*,noitemsep]
\item An \textbf{enhanced hash-based accumulator} that leverages both shared and global memory to efficiently handle long rows, and
\item \textbf{ESC accumulators} for short rows with a limited number of intermediate products.
\end{enumerate}
Here, \textit{long rows} are defined as rows with many nonzeros, typically exceeding the maximum capacity of a hash accumulator on a single streaming multiprocessor, while \textit{short rows} are rows with fewer nonzeros than the size of a warp, which is 32 on NVIDIA GPUs.

\begin{lstlisting}[style=SASS, caption={FP64 Atomic Add on Shared and Global Memory.}, label={lst:sass}, float]
        // Shared Memory FP64 Atomic Add
.L_x_8:
         LDS.64 R4, [RZ] ;
         DADD R6, R4, 1 ;
         ATOMS.CAST.SPIN.64 R6, [RZ], R4, R6 ;
         ISETP.EQ.U32.AND P0, PT, R6, 0x1, PT ;
         ISETP.EQ.U32.AND.EX P0, PT, R7, RZ, PT, P0 ;
    @!P0 BRA `(.L_x_8) ;

        // Global Memory FP64 Atomic Add
         RED.E.ADD.F64.RN.STRONG.GPU [R2.64], R8 ;
    \end{lstlisting}

\paragraph{\textbf{Enhanced Hash-Based Accumulator. }}
Hash-based accumulators must store both the column indices and the values of the output elements. 
Column indices are typically 32 bits, while values are usually 64 bits. 
Because scratchpad memory has limited capacity, rows that exceed a certain size must switch to global memory, which makes hash-based accumulators inefficient for long rows.
Our experiments show that we can store values in global memory without impacting performance under certain conditions described below, contrary to the common belief that the entire hash table must reside in scratchpad memory for fast accumulation.
This allows hash-based kernels to accumulate rows that are $3\times$ longer without significant performance penalty.
Our design is motivated by two key observations:
\begin{enumerate}[leftmargin=*,noitemsep]
\item The index access patterns and operation frequency are considerably more complex than those for value access.
The indices serve as element identifiers and must be read, compared, and swapped using atomic operations during insertion, while values are only involved in \texttt{atomicAdd} operations.
\item In practice, storing values in global memory does not significantly degrade performance because of GPU architectural features. 
Hash-based accumulators are currently limited more by memory access latency than by bandwidth.
\end{enumerate}
Prior studies show that the latency of atomic operations in global memory is only $2$–$4\times$ higher than that of shared memory atomics in general~\cite{jia_dissecting_2018}. 
Moreover, because the return value of the atomic operation is not needed in this context, the compiler uses a fire-and-forget access pattern, further reducing the impact of latency~\cite{wilt2013cuda}.
Finally, atomic operations on shared memory for \texttt{FP64} data types are not natively supported on NVIDIA GPUs and are compiled into compare-and-swap loops, reducing the advantage of shared memory placement.
Listing~\ref{lst:sass} shows the SASS code for an \texttt{atomicAdd} operation compiled for \texttt{sm80}.
For shared memory, a loop with compare-and-swap and branching instructions is required. 
For global memory, the operation compiles into a single hardware-supported SASS instruction that uses a fire-and-forget access pattern.

\paragraph{\textbf{ESC Accumulator.}}
Certain matrices exhibit extremely high sparsity, and their average number of intermediate products falls below the smallest block size (typically 64, chosen to ensure good GPU occupancy). 
In such cases, the ESC accumulator, which first expands the intermediate products and then sorts and compacts them into the final results, can be selected to take advantage of its specialization for short rows.
The ESC accumulator is particularly well suited to the upper-bound estimation workflow, as its configuration does not depend on the number of output elements. 
Instead, the number of intermediate products per row serves as the selection criterion. 
The kernel is further optimized to process multiple rows concurrently, enabling efficient accumulation for short rows.

\paragraph{\textbf{Accumulator Selection. }}
In total, three types of accumulators--hash-based, dense, and ESC--are used, each with multiple configurations for different block sizes. 
The accumulators are selected on a per-row basis to ensure proper load balance and best performance.
As in previous work, the configuration that requires the fewest resources for a given row is selected, as a smaller kernel generally helps reduce load imbalance and synchronization costs.
Dense kernels are preferred over hash-based kernels when both require the same amount of resources, except for the largest hash-based kernel, which uses our hybrid-memory design.
The ESC accumulator is selected only in the upper-bound estimation workflow.

\section{Kernel Configuration and Optimization}

In practice, achieving high performance in GPU SpGEMM requires careful kernel design. 
\ours leverages information from the analysis step to accelerate accumulation and applies architecture-aware optimizations with carefully tuned parameters.

\subsection{Assisted Kernels}

The estimation-based workflow is not chosen for matrices that would lead to high estimation and compaction costs. In these cases, symbolic computation is required, and information gathered during the analysis step can be used to accelerate symbolic accumulation.
Prior work uses the number of intermediate products as the accumulator selection metric for the symbolic step~\cite{parger_speck_2020, du_opsparse_2022}.
This approach often overestimates the required scratchpad memory, resulting in underutilization of hardware resources. 
In contrast, \ours computes the average and standard deviation of the compression ratio from sampled rows and derives a conservative estimate of the compression ratio for the entire matrix. 
The number of intermediate products for each row is then divided by this estimated ratio and used to select the accumulator.
 
\ours uses the analysis step to speed up dense accumulators in both symbolic and numeric phases, akin to assisted symbolic computation.
Dense accumulators maintain a bitmap to track non-empty elements, which is frequently updated because each insertion requires at least one bitmap access. 
\ours uses $CR$ to determine whether to query the bitmap before writing. 
Querying the bitmap increases read latency slightly but can significantly reduce write operations, which are often a bottleneck in shared memory.
Therefore, the bitmap query is enabled only when the estimated compression ratio is above a predefined threshold.

\subsection{Optimization}

For hash-based accumulators, output rows must be sorted after accumulation to comply with the CSR format.
The column index serves as the sorting key, and the corresponding accumulated value is permuted along with the key during the sorting process.
This sorting step can incur significant overhead, especially for long rows. To accelerate the sorting of key--value pairs, we use \emph{indirect sorting}~\cite{knuth1998art}.
Rather than sorting key--value pairs directly, key--\textit{ptr} pairs are generated and sorted. 
The \textit{ptr} element points to the corresponding value element and occupies fewer bits because of the small address space of on-chip memory. 
Once sorting is complete, values are written to the destination according to the sorted pointers.
This approach reduces memory movement and lowers register pressure during sorting.

In addition, we observe that most matrices have a column index range that does not require the full 32-bit representation.
Because \textit{ptr} is bounded and occupies no more than 14 bits, 
we concatenate the key and \textit{ptr} into a single 32-bit integer when possible. 
Radix sort is used, and the bits corresponding to the \textit{ptr} field are ignored during sorting~\cite{knuth1998art}.
This optimization further reduces register pressure, decreases memory traffic, and accelerates sorting.

\emph{Independent Thread Scheduling} (ITS), introduced in the Volta architecture, allows threads within a warp to diverge~\cite{nvidia_volta_tuning_guide}. 
Our profiling shows that ITS increases memory access requests and warp stall samples inside the core accumulation loop, leading to performance degradation. 
This effect can be mitigated by compiling for an earlier architecture or by manually calling \texttt{\_\_syncwarp()}. 
\ours uses the former approach.

\subsection{Parameter Configuration}
\label{subsec:param_conf}

Parameter configuration is key for high performance and, for \ours, primarily involves HyperLogLog settings, workflow selection criteria, and accumulation kernel configurations.

HLL parameters determine estimation precision. 
In \ours, we use 32 registers per HLL sketch when the input expansion ratio is less than 48, and 64 registers otherwise. 
This configuration balances estimation accuracy and computational efficiency.
An empirical analysis of estimation precision is provided in Section~\ref{subsec:eval_prec}.
The sampling ratio in the analysis step is set to $0.03$, with a minimum of $600$ and a maximum of $10,000$ sampled rows.
The threshold for enabling bitmap queries is set to 2.
The relative variance of $1/CR$ is:
$$
\frac{1}{n_{sampled}}[ \epsilon^2 + CV^2 (1+\epsilon^2)]
$$
where $CV=\sfrac{\sigma_{C}}{\mu_C}$ is the coefficient of variation of the density of output rows, and $\epsilon = \sfrac{1.04}{\sqrt{m}}$ is the relative error of the HLL estimator with $m$ registers.
Then, applying Chebyshev's inequality~\cite{saw1984chebyshev} provides a probabilistic error bound on the sampled $CR$.
For a matrix with 200,000 rows, a sampling rate of 0.03, and 64 HLL registers, the relative error is below 3\%, 6\%, and 17\% at 95\% confidence for matrices with $CV=0.5$, $1$, and $3$, corresponding to balanced, exponential, and heavily skewed output row distributions.
Given the sampled $CR$ is used for workflow selection, these error levels are acceptable even for highly skewed matrices.

Table~\ref{tab:criteria} summarizes the criteria for workflow selection. 
The upper-bound method, which skips both symbolic accumulation and HLL-based estimation, is used when the average number of intermediate products per row is below 64. 
This threshold matches the minimum number of threads per block required for full occupancy.
For the remaining matrices, the HLL-based estimation workflow is selected when both $ER$ and sampled $CR$ are greater than or equal to 8. 
The threshold for $ER$ is set to 8 because each HLL register occupies one byte, while each column index requires four bytes; at this point, HLL sketch merging incurs a memory access cost comparable to symbolic accumulation. 
The threshold for $CR$ is set to 8, chosen empirically to amortize the cost of output compaction.

For accumulation kernel configuration, \ours uses the same profile for all architectures.
The maximum shared memory allocated per SM for accumulation is capped at 128 KB, with the remaining capacity reserved for L1 cache. 
Our kernel configurations follow prior work~\cite{parger_speck_2020, du_opsparse_2022}.
In total, five normal kernels and two specialized kernels are used for both hash-based and dense accumulators. 
The largest normal kernel uses half of the available shared memory and 1024 threads to achieve full occupancy.
Each subsequent normal kernel uses half the resources of the previous one, resulting in progressively smaller accumulator sizes for shorter rows while achieving full occupancy. 
One specialized kernel uses the same resource configuration as the largest normal kernel and processes long rows using global memory. 
The other specialized kernel matches the smallest normal kernel and processes multiple short rows concurrently. 
For ESC accumulators, two configurations are provided, processing two or four rows concurrently with 64 threads.

\begin{table}[t]
  \centering
  \small
  \begin{tabular}{l p{0.6\linewidth}}
    \toprule
    Method & Selection Criteria \\
    \midrule
    Upper-bound Estimation
      & $nproducts_{\text{avg}} < 64$ \\

    HLL Estimation
      & $nproducts_{\text{avg}} \ge 64 \;\land\;
         ER \ge 8 \;\land\; CR \ge 8$ \\

    Symbolic Accumulation
      & Remaining cases \\
    \bottomrule
  \end{tabular}
  \caption{\ours's selection criteria for different SpGEMM workflows: $ER$ is the Input Expansion Ratio, and $CR$ is the (Sampled) Output Compression Ratio.
  }
  \label{tab:criteria}
\end{table}

\section{Evaluation}

This section evaluates \ours through comparisons with state-of-the-art solutions, analysis of HLL estimation and sampling accuracy, and a detailed breakdown of its core methods.

\subsection{Experiment Setup}
\label{subsec:exp_setup}

\ours is evaluated on two platforms: Perlmutter at NERSC and DeltaAI at NCSA. 
Each Perlmutter GPU node has one AMD EPYC 7763 CPU and four NVIDIA A100 GPUs with 40 GB of memory each. 
The system runs SUSE Linux Enterprise Server 15, and all experiments use CUDA 12.9. 
DeltaAI is an ARM-based supercomputer. 
Each compute node contains four NVIDIA GH200 superchips, each integrating a 72-core NVIDIA Grace CPU and an NVIDIA H100 GPU with 96 GB of memory. 
DeltaAI also runs SUSE Linux Enterprise Server 15, and CUDA 12.4 is used.

\ours is evaluated against four state-of-the-art GPU SpGEMM implementations: \textsc{spECK}~\cite{parger_speck_2020}, \textsc{opSparse}~\cite{du_opsparse_2022}, \textsc{TileSpGEMM}~\cite{niu_tilespgemm_2022}, and \textsc{HSMU-SpGEMM}~\cite{wu_hsmu-spgemm_2025}.
The code base of \textsc{MOSparse}~\cite{wang_optimizing_2025} is not open source, so we do not include a comparison with it.
For completeness, we include \textsc{cuSPARSE} in one experimental setting, although prior studies show it is generally less performant than the state of the art~\cite{gao_systematic_2023}.
For each implementation, we follow the authors' recommended guidelines to configure, compile, and run on our platforms. 
In some cases, minor code modifications are required to ensure a fair comparison.
In particular, we modify \textsc{spECK} to include output memory allocation in the reported execution time, and we adjust configurations in \textsc{HSMU-SpGEMM} to support rectangular matrices.
In our experiments, a few matrices fail the internal correctness checks of \textsc{HSMU-SpGEMM}. 
Our attempts to contact the authors were unsuccessful.

\ours and other SpGEMM approaches are evaluated on a collection of over 400 matrices, divided into two datasets for clarity: one set of square matrices and one set of rectangular matrices.
The first dataset consists of 337 square matrices from the SuiteSparse Matrix Collection. 
This dataset was originally used in \textsc{HSMU-SpGEMM}~\cite{wu_hsmu-spgemm_2025} and includes matrices from the SuiteSparse matrix collection that require at least 100 million floating-point operations (FLOPs) to compute $AA$ and $AA^T$.
FLOPs are counted as twice the number of intermediate products. 
The second dataset consists of rectangular matrices from the SuiteSparse Matrix Collection, filtered to have no more than 10 billion nonzeros and at least 100 million FLOPs for computing $AA^T$, resulting in 64 matrices. 
The matrices from the two datasets cover a wide range of applications and display diverse structural properties, such as matrices that produce dense output and symmetric matrices.
For the square dataset, the evaluation focuses on computing $AA$. 
For the rectangular dataset, we evaluate the computation of $AA^T$.

For each test, 10 warm-up runs are followed by 10 measured runs, and the average runtime of the measured runs is reported. 
A timeout of 30 seconds is imposed for all runs. 
If a tool fails on a matrix due to a runtime error or timeout, its runtime is replaced with the slowest valid runtime observed for that matrix across all tools. 
This penalizes failures, prevents artificially inflated performance, and ensures comparable results.

\subsection{Overall Performance}
\label{subsec:overall_perf}

\begin{table}[t]
  \centering
  \begin{tabular}{l l c c c c}
    \toprule
    GPU (\textit{dataset)} & Tool & \#best & \#2nd best & \#fail & GFLOPS  \\
    \midrule
    \multirow{6}{*}{\shortstack{A100 (\textit{square})}}
    & \textsc{cuSparse} & 0 & 0 & 82 & 3.39 \\
    & \textsc{spECK} & 22 & 185 & 0 & 46.2 \\
    & \textsc{opSparse} & 16 & 86 & 20 & 24.2 \\
    & \textsc{Tile} & 5 & 9 & 82 & 18.1 \\
    & \textsc{HSMU} & 0 & 19 & 0 & 32.1 \\
    & \textbf{\ours} & \textbf{294} & 38 & 0 & \textbf{63.7} \\
    \midrule
    \multirow{3}{*}{\shortstack{H100 (\textit{square})}}
    & \textsc{spECK} & 21 & 204 & 0 & 66.2 \\
    & \textsc{opSparse} & 4 & 110 & 18 & 40.1 \\
    & \textbf{\ours} & \textbf{312} & 23 & 0 & \textbf{108.0} \\
    \midrule
    \multirow{4}{*}{\shortstack{A100 (\textit{rect.})}}
    & \textsc{spECK} & 7 & 39 & 0 & 27.3 \\
    & \textsc{opSparse} & 15 & 3 & 4 & 10.8 \\
    & \textsc{HSMU} & 0 & 6 & 15 & 15.3 \\
    & \textbf{\ours} & \textbf{42} & 16 & 1 & \textbf{36.3} \\
    \midrule
    \multirow{3}{*}{\shortstack{H100 (\textit{rect.})}}
    & \textsc{spECK} & 8 & 38 & 0 & 42.6 \\
    & \textsc{opSparse} & 8 & 12 & 4 & 17.1 \\
    & \textbf{\ours} & \textbf{48} & 14 & 0 & \textbf{60.7} \\
    
    \bottomrule
  \end{tabular}
\caption{Comparison of SpGEMM approaches on different datasets and architectures. \
The \textit{square} dataset includes 337 matrices. \
The \textit{rectangular} dataset includes 64 matrices. \
The average GFLOPS is the geometric mean across matrices. 
}
\label{tab:comp_table}
\end{table}

\begin{figure}[t]
  \centering
  \includegraphics[width=\columnwidth]{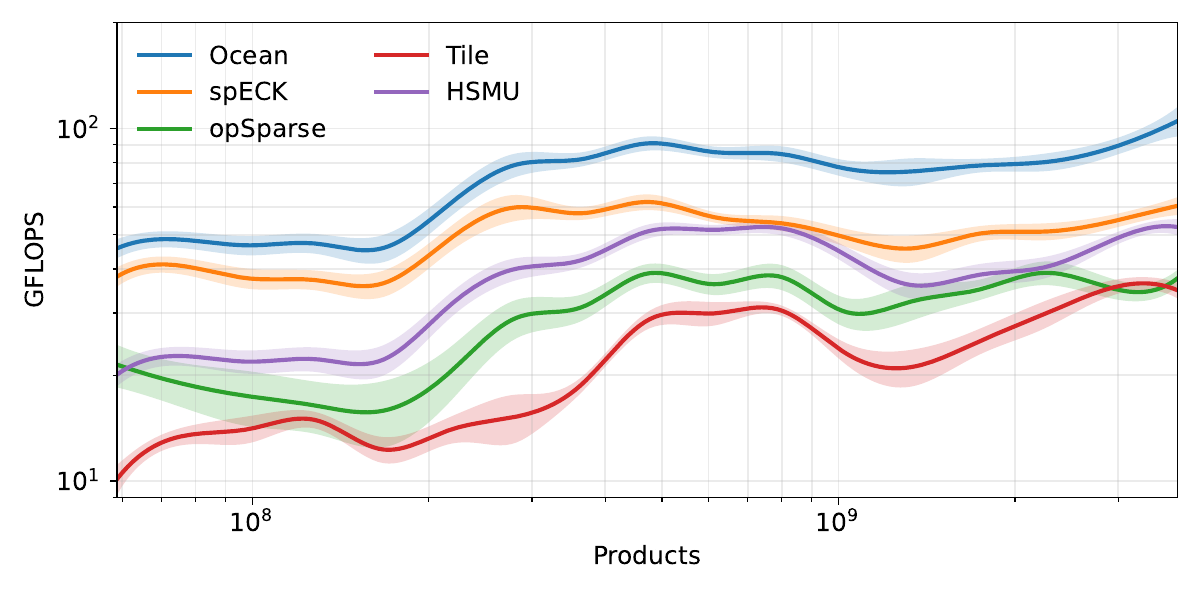}
  \caption{The smoothed GFLOPS achieved over the square matrices dataset on the A100. Ordered by the number of intermediate products. The line shade indicates the deviation.}
  \label{fig:comp_table_trend}
\end{figure}

\begin{figure*}[t]
\setlength{\abovecaptionskip}{0pt}
\setlength{\belowcaptionskip}{0pt}
\centering
\begin{subfigure}{0.5\textwidth}
  \centering
  \includegraphics[width=\linewidth]{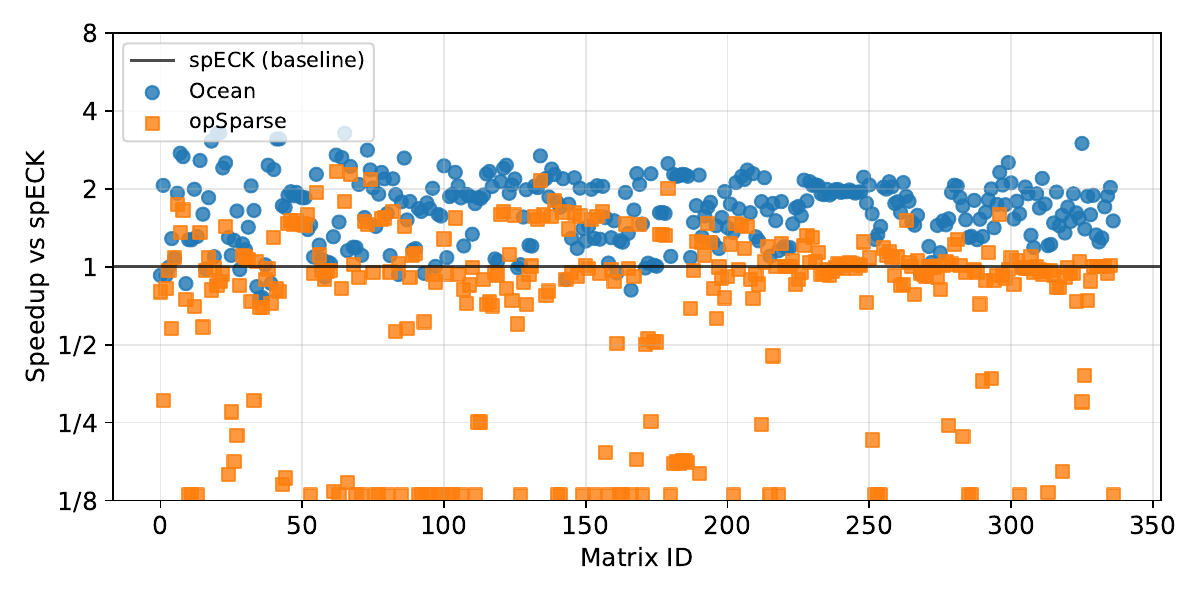}
  \label{fig:sub1}
\end{subfigure}%
\begin{subfigure}{0.5\textwidth}
  \centering
  \includegraphics[width=\linewidth]{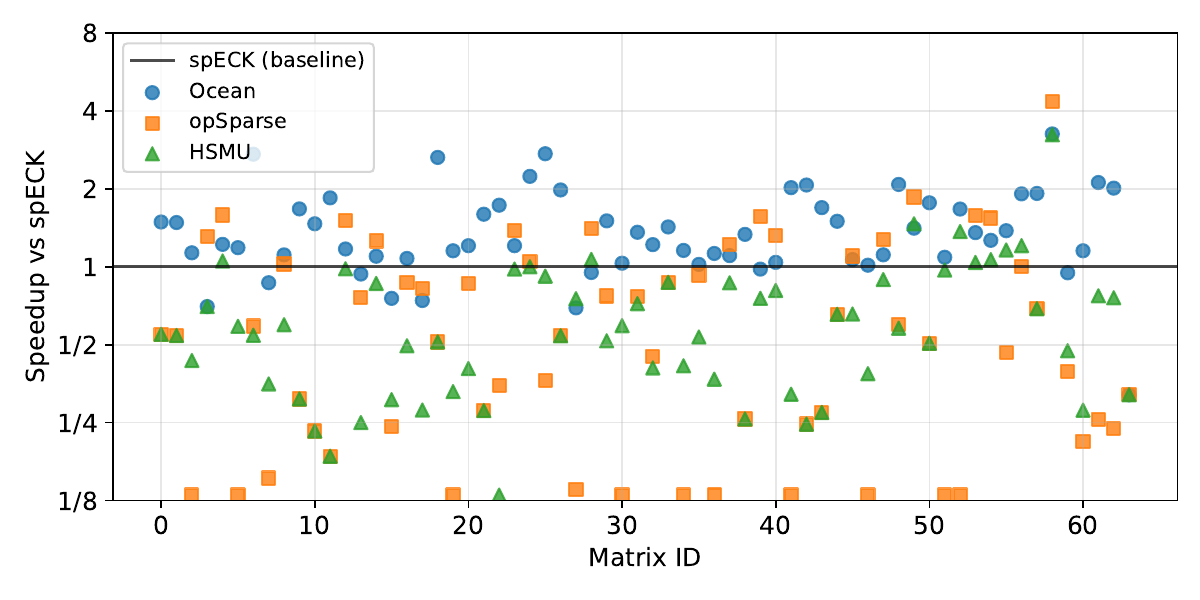}
  \label{fig:sub2}
\end{subfigure}
\caption{Relative speedup of SpGEMM solutions compared to \textsc{spECK}. 
The speedups below $1/8$ are clipped to $1/8$ in the plot. Results for square matrices (left) are shown on H100, and results for rectangular matrices (right) are shown on A100.
}
\label{fig:comp_scatter}
\end{figure*}

The first section of Table~\ref{tab:comp_table} summarizes the overall performance statistics for the square-matrix dataset on NVIDIA A100 GPUs. 
Figure~\ref{fig:comp_table_trend} illustrates the corresponding performance trends, with the number of intermediate products on the x-axis and the average achieved GFLOPS on the y-axis.
Out of the 337 matrices, \ours achieves the best performance on 86\% of the matrices and ranks second on an additional 11\%.
\textsc{cuSparse}, \textsc{spECK}, \textsc{opSparse}, \textsc{TileSpGEMM}, and \textsc{HSMU-SpGEMM} achieve the best performance on 0\%, 7.8\%, 3.9\%, 1.7\%, and 0\% matrices, respectively.
\ours achieves a geometric mean performance of 63.23 GFLOPS. 
On average, \ours outperforms \textsc{cuSparse}, \textsc{spECK}, \textsc{opSparse}, \textsc{TileSpGEMM}, and \textsc{HSMU-SpGEMM} by a factor of $18.8\times$, $1.4\times$, $2.6\times$, $3.5\times$, and $2\times$, respectively.

The trend plot illustrates the performance of different SpGEMM approaches across matrices with varying numbers of intermediate products. 
As shown in the figure, \ours consistently achieves the highest performance across the entire dataset, and its performance advantage increases as the number of intermediate products grows. 
This trend can be attributed to two main factors:
\begin{enumerate}[leftmargin=*,noitemsep]
\item The relative overhead of the analysis components decreases as overall computation increases;
\item The estimation-based workflow and enhanced hash-based accumulators are effective at handling longer rows, making \ours more efficient for larger matrices.
\end{enumerate}

Of the six evaluated implementations, \textsc{spECK} achieves the second-best overall performance. 
Its lightweight analysis, load-balancing strategy, and hybrid use of hash and dense accumulators enable consistently strong performance across a wide range of matrices.
In addition, \textsc{spECK} disables global load balancing when the analysis results are uniform across all rows. 
This reduces overhead for certain matrices and allows \textsc{spECK} to outperform \ours in these cases. 
\textsc{HSMU-SpGEMM} ranks third in overall performance. Our results suggest that the binary search-based design of its numeric accumulator has higher complexity than hash-based accumulators, which outweighs its greater utilization of shared memory. 
\textsc{opSparse} generally ranks fourth in performance.
It outperforms \ours and \textsc{spECK} on a small number of matrices, most of which are extremely sparse, due to its overhead optimizations and specialized design for the smallest accumulator kernel.
However, its design prevents it from achieving consistent performance across matrices, a limitation also reflected in the high variance in performance. 
\textsc{TileSpGEMM} exhibits the second-lowest overall performance, although it achieves the best results on six matrices. 
Its tile-based computation strategy is effective primarily for matrices with block-sparse structure, which limits its general applicability.

The square dataset is also evaluated on DeltaAI, equipped with NVIDIA H100 GPUs. 
Because DeltaAI is an ARM-based system, \textsc{TileSpGEMM} and \textsc{HSMU-SpGEMM} fail to compile as they rely on x86 intrinsics and are therefore excluded from this evaluation.
The results are presented in the second section of Table~\ref{tab:comp_table}.
Figure~\ref{fig:comp_scatter} (left) further illustrates the relative speedup of \ours and \textsc{opSparse}, using \textsc{spECK} as the baseline. 
\textsc{spECK} is chosen as the baseline because it successfully runs on all matrices and exhibits stable performance. 
\ours achieves consistent speedup across the dataset, being the best-performing solution for 92\% of the matrices, with average speedups of $1.6\times$ over \textsc{spECK} and $2.7\times$ over \textsc{opSparse}.

In addition, we evaluate the rectangular dataset on NVIDIA A100 GPUs to demonstrate the generality of \ours. 
\textsc{TileSpGEMM} does not support rectangular matrices and is therefore excluded from this evaluation. 
Figure~\ref{fig:comp_scatter} (right) shows the relative speedup of the four remaining SpGEMM implementations.
\ours demonstrates consistently strong performance on this dataset, achieving the best performance on 42 matrices and the second-best performance on 18 matrices. 
On average, \ours achieves speedups of $1.3\times$, $3.4\times$, and $2.4\times$ over \textsc{spECK}, \textsc{opSparse}, and \textsc{HSMU-SpGEMM}, respectively. 
The evaluation on H100 is provided in Table~\ref{tab:comp_table}.

On the A100, \ours fails on one matrix, \texttt{JP}, due to memory constraints.
The estimation-based approaches inherently incur a peak memory usage of about $2.2\times$ because of GPU memory management limitations.
Specifically, results produced by the numeric step are not stored in consecutive memory locations and must be compacted to satisfy the CSR format.
A small auxiliary buffer can compact non-consecutive outputs back into the original array, but the array remains over-allocated because in-place shrinking is not supported on GPUs.
In situations where GPU memory is limited, this issue can potentially be mitigated by falling back to the symbolic step if insufficient memory is detected.
The input expansion ratio and the sampled output compression ratio can be used to estimate the memory footprint of the estimation-based workflow.

\vspace{-.5em}
\subsection{Estimation Precision}
\label{subsec:eval_prec}

\begin{figure}[t]
  \centering
  \includegraphics[width=\columnwidth]{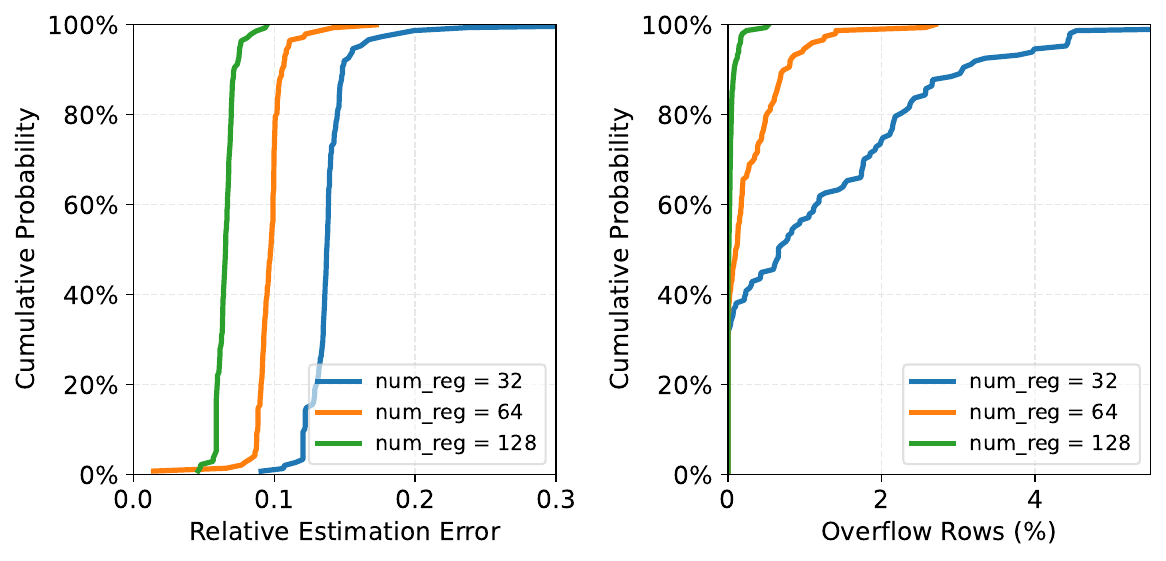}
  \caption{CDF of the average relative estimation error of HLL Estimators (left) and ratio of overflow rows (right) under different number of HLL registers on the Square dataset.}
  \label{fig:cdf}
\end{figure}

To evaluate the precision of HyperLogLog under different register counts, we conduct experiments on the square dataset, excluding matrices that do not use the HLL-based estimation approach. 
The number of registers per HLL sketch is set to 32, 64, and 128, and the accuracy of per-row nonzero estimation is measured. 
Figure~\ref{fig:cdf} (left) shows the cumulative distribution function (CDF) of the average relative estimation error. 
For each matrix, the relative estimation error is calculated as the arithmetic mean of the per-row relative errors. 
The average errors across all matrices with 32, 64, and 128 registers are 0.13, 0.10, and 0.07, respectively, with corresponding RMS averages of the standard deviation of 0.17, 0.15, and 0.14. 
These results indicate that increasing HLL precision consistently reduces estimation errors.
They also show that HLL can provide sufficient accuracy for per-row nonzero estimation with a reasonable number of registers.
Our empirical results show that fixing the register count at 32 or 64 causes an average overall slowdown of 1.4\% and 1\%, respectively, compared to the dynamic configuration described in Section~\ref{subsec:param_conf}.

To compare the precision of \ours's HLL estimator with Cohen's estimator~\cite{cohen1996optimizing}, both estimators are configured to use the same amount of memory per output row, as estimation kernels are memory-bound. 
Cohen's estimator has a memory access pattern similar to HLL on matrix $\A$ but requires substantially more memory accesses on matrix $\B$.
Therefore, this configuration favors Cohen's estimator. 
Using 64 bytes per output row, HLL achieves a smaller average relative error for per-row estimation on all 148 matrices in the HLL estimation-based workflow, with an average error $2.1\times$ smaller than Cohen's. 
Even when Cohen's estimator is given $4\times$ more memory, HLL still outperforms it on 116 of the 148 matrices, consistent with theoretical expectations. 
These results show that HLL provides more accurate per-row output size predictions for SpGEMM than existing estimation approaches.

In addition, we evaluate the fraction of overflow rows using HLL-based estimation. 
The HLL estimates are treated the same as symbolic results for binning.
Each estimate is first multiplied by a coefficient (1.5 by default) and then rounded up to the nearest accumulator size. 
A row is considered overflowing when the actual number of output elements exceeds $80\%$ and $100\%$ of the allocated capacity for hash-based accumulators and dense accumulators, respectively. 
In this setting, the average overflow ratios are 1.2\%, 0.3\%, and well below 0.1\% for 32, 64, and 128 registers, respectively.
The corresponding maximum overflow ratios are 7\%, 2.7\%, and 0.5\%.
Figure~\ref{fig:cdf} (right) shows the cumulative distribution of overflow ratios. 
These results indicate that overflows caused by estimation are rare and have limited impact on numeric accumulation. 
To further mitigate overflows at lower precision, we increase the expansion coefficient to 2.0 when using 32 registers.

The sampling strategy proposed in Section~\ref{subsec:est_wkfl} enables accurate estimation of the output compression ratio $CR$.
On the square dataset, the average relative sampling errors are 0.05, 0.04, and 0.03 for 32, 64, and 128 registers, respectively.
For the three register configurations, only 2, 1, and 1 matrices are assigned to different workflow-selection categories when comparing the sampled $CR$ with the ground-truth $CR$.
These error levels are sufficient for reliable workflow selection.
The sampling rate of 0.03 is chosen empirically to balance accuracy and overhead: reducing it to 0.01 causes a 2.1\% average slowdown, as the less accurate $CR$ estimate leads to suboptimal workflow selection despite benefiting 21 individual matrices; increasing it to 0.05 causes a 3.5\% slowdown due to the higher sampling cost.

\subsection{Contribution Breakdown}
\label{subsec:cont_breakdown}

\begin{table*}[t]
  \centering
  \begin{tabular}{cl ccc ccc cc}
    \toprule
    & \multicolumn{1}{c}{}
    & \multicolumn{3}{c}{Symbolic-based Speedup}
    & \multicolumn{3}{c}{Estimation-based Speedup}
    & \multicolumn{2}{c}{Overall Avg.} \\
    \cmidrule(lr){3-5} \cmidrule(lr){6-8} \cmidrule(lr){9-10}
    Version & Method & Avg. & Min. & Max. & Avg. & Min. & Max. & Speedup & GFLOPS  \\
    \midrule
    V1 & Baseline  & - & - & - & - & - & - & - & 50.8\\
    V2 & (V1)+E    & - & - & - & $1.30\times$ & $0.95\times$ & $2.77\times$ & $1.14\times$  & 58.3 \\
    V3 & (V2)+AS   & $1.04\times$ & $0.96\times$ & $1.65\times$ & $1.03\times$ & $0.99\times$ & $2.19\times$ & $1.03\times$ & 60.2 \\
    V4 & (V3)+HA   & $1.07\times$ & $0.93\times$ & $1.63\times$ & $1.04\times$ & $0.55\times$ & $2.53\times$ & $1.06\times$ & 63.7 \\
    \midrule
    \multicolumn{2}{c}{\emph{Overall}} & $1.12\times$ & $0.92\times$ & $1.66\times$ & $1.40\times$ & $0.99\times$ & $3.77\times$ & $1.25 \times$ & 63.7\\ 
    \bottomrule
  \end{tabular}
  \caption{
  Comparison of \ours with different optimization strategies.
  }
  \label{tab:ablation}
\end{table*}

\begin{figure}[t]
  \centering
  \includegraphics[width=\columnwidth]{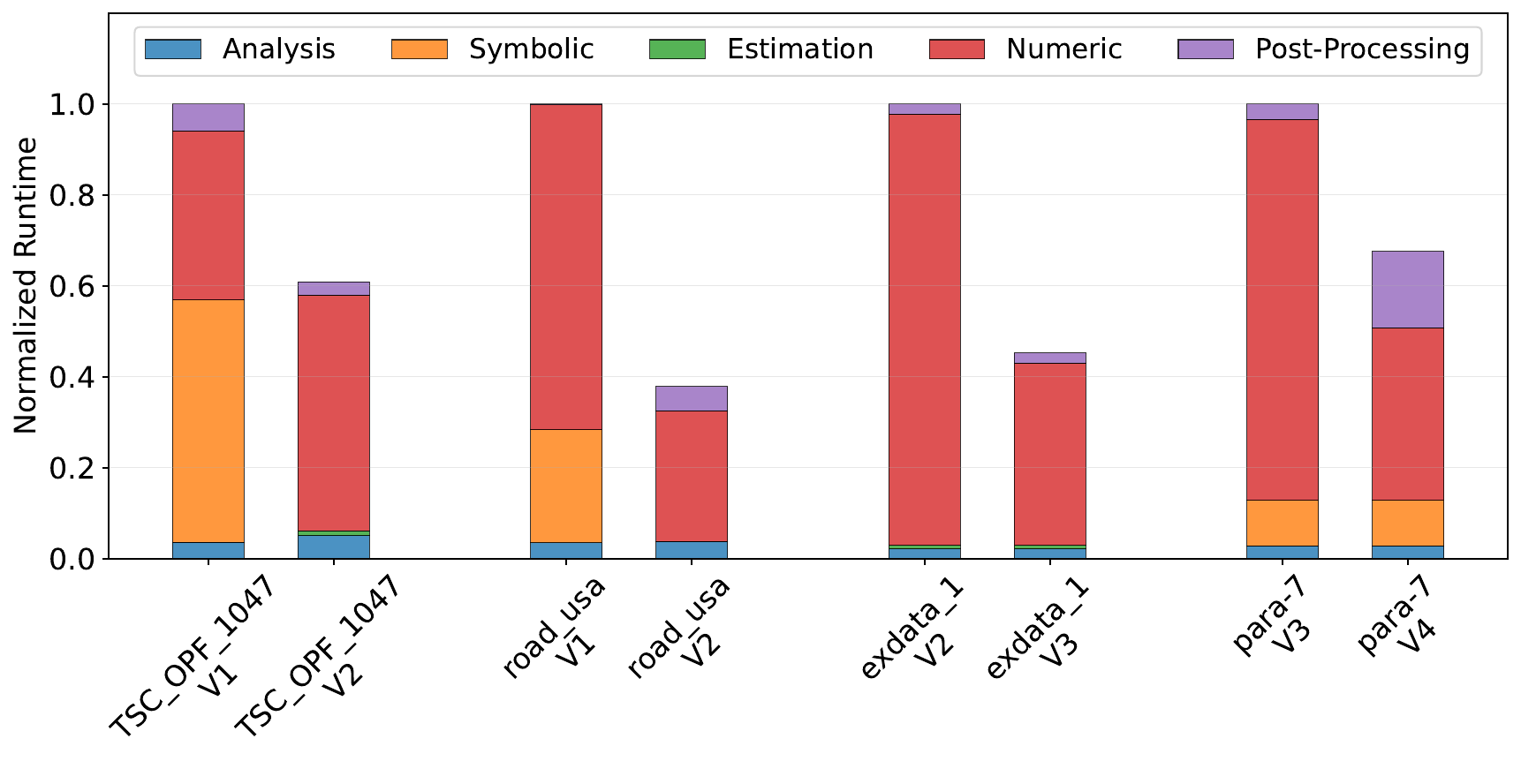}
  \caption{
  Runtime breakdown of individual components across different versions. 
  Runtimes are normalized to the first version for each matrix for clarity.
  }
  \label{fig:component_runtime}
\end{figure}

Finally, we assess \ours's core innovation through an ablation study on the square dataset using an A100, evaluating four incremental versions of \ours.
The baseline (V1) disables the estimation-based workflow, the assisted kernels, and the hybrid accumulators (including the enhanced hash-based accumulator and ESC accumulators). 
V2 introduces the estimation-based workflow (labeled E), V3 adds the assisted kernels (labeled AS), and V4 incorporates hybrid accumulator support (labeled HA). 
Table~\ref{tab:ablation} summarizes the results. Each version is compared with its immediate predecessor to report incremental speedup, while the final row labeled \emph{Overall} compares V4 with V1. 
Results are reported separately for matrices that use symbolic-based prediction and those that use estimation. 
Of the 337 matrices, 148 are assigned to the HLL-based estimation workflow according to the criteria in Table~\ref{tab:criteria}, 26 use upper-bound estimation, and the remaining 163 use precise prediction.
\ours's performance is generally consistent across input matrices of varying density throughout the dataset.

The estimation-based workflow provides an average speedup of $1.3\times$ for the selected matrices by significantly reducing the cost of the symbolic step.
The speedup is greater for matrices with density in $[10^{-2}, 10^{-1})$ and for those below $10^{-5}$.
The maximum speedups are observed on matrices \texttt{TSC\_OPF\_1047} for HLL-based estimation and \texttt{road\_usa} for the upper-bound method, respectively.
Figure~\ref{fig:component_runtime} further illustrates the breakdown of runtime for individual stages.
For \texttt{TSC\_OPF\_1047}, HLL-based estimation substantially reduces the cost of size prediction compared to the symbolic step, but slightly increases numeric kernel time due to suboptimal binning. For \texttt{road\_usa}, the upper-bound method eliminates the cost of output prediction and improves numeric performance through more effective binning guided by the upper bound.

To validate the cost-prediction model and the accuracy of workflow selection, we conduct an additional evaluation using $V4$, forcing matrices to use either the estimation-based workflow or the symbolic-based workflow.
Of the 163 matrices that initially selected the symbolic-based workflow, only $10.8\%$ benefit from switching to the estimation-based workflow with a speedup of at least $1.05\times$.
In contrast, among the 174 matrices that originally selected the estimation-based workflow, $92\%$ achieve a speedup greater than $1.05\times$ compared to reverting to the symbolic-based workflow.
These results demonstrate that our cost-prediction model successfully selects the optimal workflow for most matrices.

The assisted kernels provide an additional average speedup of $1.03\times$, with a maximum speedup of $2.19\times$ on the matrix \texttt{exdata\_1}. 
This matrix produces many dense output rows with a very high compression ratio, where frequent updates to the bitmap cause substantial memory traffic. 
As shown in Figure~\ref{fig:component_runtime}, the assisted kernels reduce the numeric step runtime by more than 50\%.
Overall, the performance gains from assisted kernels, particularly the assisted symbolic step, are modest.
Our results show that, although an estimated compression ratio allows for allocating less shared memory and selecting smaller kernels, this does not always improve performance. 
Kernel efficiency depends on multiple factors, including intermediate products and load balancing. 
These results indicate that smaller kernel sizes do not inherently yield better performance, and additional metrics should guide kernel assignment.

The hybrid accumulator provides an average speedup of $1.06\times$. 
The enhanced hash-based accumulator benefits three times as many matrices as the ESC accumulator, though each method yields an average speedup of $1.2\times$ on its respective set of affected matrices.
The largest performance improvement is observed on matrix \texttt{SiO2}, which has a small number of scattered rows requiring multiple dense-accumulator iterations, leading to a tail effect.
Figure~\ref{fig:component_runtime} illustrates the runtime breakdown for another matrix, \texttt{para-7}, which also benefits from the hybrid accumulator.
Rows that would otherwise require multiple dense accumulator iterations are instead handled by the enhanced hash-based accumulator. 
The time spent on post-processing, including output sorting, increases, but overall performance improves, achieving a $1.6\times$ speedup.
A notable exception is \texttt{torso1}, which experiences a slowdown of $0.55\times$ and is the only matrix with a slowdown greater than $0.9\times$. 
For \texttt{torso1}, 1,215 out of 116,158 rows are assigned to the enhanced hash-based accumulator. 
These rows generate a large number of intermediate products, resulting in high global memory traffic and performance degradation. 
The enhanced hash-based accumulator effectively hides memory access latency; however, adding a mechanism to detect and avoid throughput-bound cases would further improve its robustness.

Finally, we analyze the runtimes of different components.
For \ours, analysis takes 7\% of runtime, with HLL sampling just 2\%, confirming the analysis is lightweight.
For matrices processed with HLL-based estimation, the estimation step takes an average of 4\% of the runtime, and post-processing accounts for about 8\%.
In contrast, with the baseline version, the symbolic step for such matrices consumes 30\% of the runtime, while post-processing accounts for about 4\%.
As a reference, the symbolic step takes an average of 28\% of the total runtime in \textsc{spECK}. 
These results show that, although memory compaction cost increases, estimation incurs much less overhead than symbolic computation, resulting in a net performance gain.

\section{Related Work}

\textsc{cuSparse}~\cite{cusparse} is one of the earliest SpGEMM solutions to use a two-pass (symbolic and numeric) approach. 
It uses hash-based accumulators for accumulation.
\textsc{spECK}~\cite{parger_speck_2020} follows the classic two-pass design. It uses both hash-based and dense accumulators. A lightweight analysis is introduced to dynamically select among different accumulators and algorithms. Its design aims to deliver robust performance across a wide range of input matrices. \textsc{opSparse}~\cite{du_opsparse_2022} also follows the two-pass approach and proposes multiple low-level, architecture-specific optimizations.
\textsc{TileSpGEMM}~\cite{niu_tilespgemm_2022} uses tiles as the minimum unit of computation. 
This novel computation strategy eliminates the need for symbolic computation, but a similar preprocessing step is still required to determine the output tile pattern.
\textsc{HSMU-SpGEMM}~\cite{wu_hsmu-spgemm_2025} replaces hash-based accumulation with binary search on a pre-generated column-index array, eliminating the need for a load factor and improving shared memory utilization by about $1.5\times$. \textsc{HSMU-SpGEMM} also follows the two-step design, but the symbolic phase is modified to generate the column-index array.
\textsc{MOSparse}~\cite{wang_optimizing_2025} uses a LightGBM-based model to select between the two-pass ``precise'' method and the symbolic-free ``upper-bound'' method, while also incorporating architecture-specific optimizations and parameter tuning. 
The code is not publicly available.
\textsc{AC-SpGEMM}~\cite{winter2019adaptive} emphasizes its use of ESC accumulators with adaptive chunking. 
It performs well on very sparse matrices but is generally slower than newer approaches such as \textsc{spECK}~\cite{parger_speck_2020, gao_systematic_2023}. 
Therefore, we focus on more recent GPU SpGEMM approaches.
Orthogonal to single-GPU approaches, recent work has addressed distributed multi-GPU SpGEMM using RDMA-based communication and hierarchy-aware decomposition~\cite{brock2024rdma, trident}. 
In particular, Bellavita et al. introduced trident partitioning, a 2D-1D hybrid scheme that reduces internode communication by leveraging fast intranode GPU interconnect~\cite{trident}.
These distributed approaches rely on a local single-GPU SpGEMM kernel; \textsc{Ocean} could serve as such a kernel to improve end-to-end performance.

Overall, existing GPU SpGEMM solutions prioritize accumulation in shared memory. 
If shared memory is insufficient, they fall back to global memory. 
To our knowledge, no previous work has proposed hash-based accumulators that utilize both shared and global memory, or combined hash-based, dense, and ESC accumulators within a single design. 
Prior GPU SpGEMM solutions do not use estimation-based techniques to replace per-row size prediction. 
\textsc{MOSparse} sometimes avoids symbolic computation using the ``upper-bound'' method, but it lacks probabilistic estimation and faces two limitations: (i) potentially high GPU memory usage depending on output compression, and (ii) reliance on training data, which may not generalize to unseen matrices.

Prior work has explored output size estimation for SpGEMM, but they do not target GPUs. 
Cohen et al.~\cite{cohen1996optimizing, cohen_size-estimation_1997} formulated the problem as a graph problem and applied Monte Carlo-based algorithms for fast estimation. 
Their approach can be applied to a chain of sparse matrix multiplications.
More recently, Amossen et al.~\cite{amossen_better_2014} improved the algorithm using pairwise independent hash functions. 
Du et al.~\cite{du_predicting_2022} proposed a sampling-based estimation method for CPUs.
The latter two approaches are not directly applicable to fast GPU-based SpGEMM for two main reasons: (i) their estimation procedures cannot be efficiently parallelized on GPUs, and (ii) they do not provide direct per-row output estimation, which is critical on GPUs due to scratchpad memory constraints.
Dividing the per-row intermediate product count by a global compression ratio gives only a rough estimate and can lead to significant errors in computations with non-uniform compression ratios.
Cohen's estimator supports per-row output estimation and has access patterns for matrix $\A$ similar to \ours's HLL approach. 
It, however, requires multiple rounds of random access to the nodes corresponding to columns of matrix $\B$, making it less efficient than \ours. 
In addition, it does not support fast sampling or $CR$ estimation.
\section{Conclusion}

SpGEMM is a critical kernel for simulations, graph analytics, and machine learning, yet most GPU implementations rely on a two-pass variant of Gustavson’s algorithm to optimize accumulation. 
This work questions the necessity of the initial GPU pass for precise size prediction and instead proposes a fast, estimation-driven approach.
In particular, we introduce HyperLogLog as a lightweight estimator for SpGEMM, demonstrating that it offers sufficient accuracy and is highly parallelizable and efficient on GPUs.
In addition, we design a complete workflow, \ours, that incorporates sampling and method selection to enable effective integration of HyperLogLog into GPU SpGEMM.
Building on this estimation-centered design, we optimize accumulation kernels using sampled information and further propose a hybrid accumulator scheme.
Our extensive performance comparisons and ablation studies across multiple versions of \ours validate the effectiveness of our design choices. 
Overall, \ours consistently outperforms existing implementations on more than 400 matrices.

Key directions for future work include predicting accumulation kernel performance across configurations, since shared memory requirements alone provide only a lower bound on efficiency.
Understanding the structure of input matrices and adopting a more fine-grained workflow selection can also lead to higher performance.
Beyond SpGEMM, HyperLogLog also shows promise for tasks such as row reordering and workload characterization. 
Output structure prediction is unique to SpGEMM, but HLL could also be applied to other sparse linear algebra primitives and tasks, such as guiding row reordering or characterizing workloads for SpMM.

\begin{acks}
This research used resources from the National Energy Research Scientific Computing Center, a DOE Office of Science User Facility supported by the Office of Science of the U.S. Department of Energy under Contract No. DE-AC02-05CH11231, using NERSC award ASCR-ERCAP0030076. 
This work used DeltaAI at the National Center for Supercomputing Applications (NCSA) through allocation CIS251351 from the Advanced Cyberinfrastructure Coordination Ecosystem: Services \& Support (ACCESS) program, which is supported by U.S. National Science Foundation grants \#2138259, \#2138286, \#2138307, \#2137603, and \#2138296.
This material is based upon work supported by the National Science Foundation under Grant IIS-2435801. The authors disclose that generative AI and editing assistants were used for grammar checking and to improve the clarity of the writing.
The authors thank Julian Bellavita for reading the manuscript and providing feedback, and Qingyao Sun for contributing to the error analysis of $CR$ sampling.
\end{acks}

\bibliographystyle{ACM-Reference-Format}
\bibliography{ref}


\appendix

\end{document}